\documentclass[12pt, onecolumn]{IEEEtran}
\usepackage{latexsym, epsfig}                                                          % if you need a4paper

\title{\Large \bf On Energy Efficient Hierarchical Cross-Layer Design: \\ Joint Power Control and Routing for Ad Hoc Networks\footnote{This work was presented in part at the 42nd IEEE Conference on Decision and Control, Maui, HI, December 2003 ; This research was supported by the National Science Foundation under Grants ANI-03038807 and CCR-02-05214, and by the New Jersey Center for Pervasive Information Technology. 
}}
\author{Cristina Comaniciu  \hspace{2.3cm}H. Vincent Poor\\ 
Stevens Institute of Technology  \hspace{0.7cm} Princeton University \\
e-mail: {ccomanic@stevens.edu  \hspace{1.5cm}poor@princeton.edu}}

\newcommand{\p}{{\bf p} }
\newcommand{\w}{{\bf w} }
\newcommand{\Pp}{{\bf P} }
\newcommand{\ci}{\mbox{{\bf c$_i$}}}
\newcommand{\si}{\mbox{{\bf s$_i$}}}
\newcommand{\sj}{\mbox{{\bf s$_j$}}}
\newcommand{\A}{\mbox{{\bf  A}}}

\newtheorem{theorem}{Theorem}
\newtheorem{prop}{Proposition}
\begin{document}

\maketitle
\thispagestyle{empty}

%%%%%%%%%%%%%%%%%%%%%%%%%%%%%%%%%%%%%%%%%%%%%%%%%%%%%%%%%%%%%%%%%%%%%%%%%%%%%%
%\renewcommand{\baselinestretch}{2.0}
\begin{abstract}
In this paper, a hierarchical cross-layer design approach is proposed to
increase energy efficiency in ad hoc networks through joint adaptation 
of nodes'
transmitting powers and route selection. The design maintains
the advantages of the classic OSI model, while accounting for
the cross-coupling between layers, through information sharing.
The proposed joint power control and routing algorithm is shown to 
 increase significantly the overall energy efficiency of the network, at 
the expense of a moderate increase in complexity. 
Performance enhancement of the joint design using multiuser detection 
is also investigated, and it is shown that the use of multiuser detection can
increase the capacity of the ad hoc network significantly for a given level of energy consumption.

\end{abstract}

\vspace{.5cm}

\hspace{1cm} {\small \bf Keywords: ad hoc networks, power control, routing, cross-layer}
\vspace{1cm}

%%%%%%%%%%%%%%%%%%%%%%%%%%%%%%%%%%%%%%%%%%%%%%%%%%%%%%%%%%%%%%%%%%%%%%%%%%%%%%%%
\pagebreak
\setcounter{page}{1}
\section{Introduction}

A mobile ad hoc network consists of a group of mobile nodes that spontaneously 
form temporary networks without the aid of a fixed infrastructure or 
centralized management. Ad-hoc networks rely on peer to peer communication, 
where any source-destination pair of nodes can either communicate directly
or by using intermediate nodes to relay the traffic. The communication routes are
determined by the routing protocol, which finds the best possible routes 
according to some specified cost criterion. Since, in general, many ad hoc 
networks will consist of small terminals with limited battery lifetime, 
routing protocols using energy related cost criteria have recently  been
investigated in the literature (e.g. \cite{r1},\cite{r2},\cite{r3},\cite{admud}).

Aside from ``energy aware routing'', other interference management techniques 
have the potential of improving the system performance, with a direct effect 
on increasing the network lifetime. For example, joint power control and 
scheduling have been proposed in \cite{powsch}, and power aware routing for 
networks using blind multiuser receivers has been analyzed in \cite{admud}. 
The benefits of power control for wireless networks have been shown in numerous works
(see for example \cite{pow1, pow2, pow3, meshk1}), but only recently have its interaction with 
``energy aware routing'' begun to be addressed \cite{CDC03, zorzi1, kish, comaniciu_book}.

A power aware routing protocol design relies on the current power assignments 
at the terminals, and in turn, optimal power assignment depends on the 
current network topology, which is determined by routing. It is apparent that 
there is a strong cross-coupling between power control and routing, due to 
the fact that they are both affected by, and act upon, the interference 
level and the interference distribution in the network. Given this strong
coupling between layers, we expect that cross-layer interference management
algorithms will outperform independently designed algorithms associated with
various layers of the protocol stack \cite{gold_adhoc}.
On the other hand, a concern associated with crossing the boundaries between
layers is that many of the core advantages of the OSI model, such as easy 
debugging and flexibility, easy upgrading, and hierarchical time scale adaptation, may be lost \cite{caut}.

As a tradeoff between the pros and cons of cross-layer design, we propose
a hierarchical cross-layer design framework, in which the adaptation protocols 
at different layers of the protocol stack are independently designed (e.g. 
power control, at the physical layer, and routing, at the network layer), 
while sharing coupling information across layers. Based on this framework
we propose and analyze a joint power control and 
routing algorithm for Code Division Multiple Access (CDMA) ad hoc networks. 
We then extend this algorithm to include multiuser detection, for a further 
increase in network performance. 

The paper is organized as follows: we first present the hierarchical cross-layer design framework in Section II. We then propose a joint power control and 
routing algorithm in Section III, and we add multiuser detection capabilities
for the physical layer in Section IV. Finally, Section V presents the conclusions.

\section{Hierarchical Cross-Layer Design Framework}

As we have already mentioned, a tight coupling exists between different 
interference management algorithms implemented at various layers of the 
protocol stack. In this paper we concentrate mainly on interactions between
the physical and the network layer, namely, we consider power control and
receiver adaptation algorithms at the physical layer, and energy aware routing
at the network layer. While power control and multiuser detection are 
traditional interference management techniques, energy aware routing can 
also be seen as an effective interference management tool, as seeking low 
energy routes may lead to a better interference distribution in the network.

Given the tight cross-coupling among these techniques, it becomes apparent 
that a cross-layer solution that jointly optimizes interference management 
algorithms across layers is desirable.
On the other hand, the OSI classical layered architecture has a number of 
advantages such as deployment flexibility and upgradeability, easy debugging, 
and last but not least, an inherent reduced network overhead by implementing 
adaptability at different time scales. More specifically,
fast adaptation can be done locally by the physical layer, while large scale 
events can be handled by changes in routing, which implies at least local 
neighborhood information updates.

\begin{figure}[ht]
\centerline{
\epsfxsize=2.5 in\epsffile{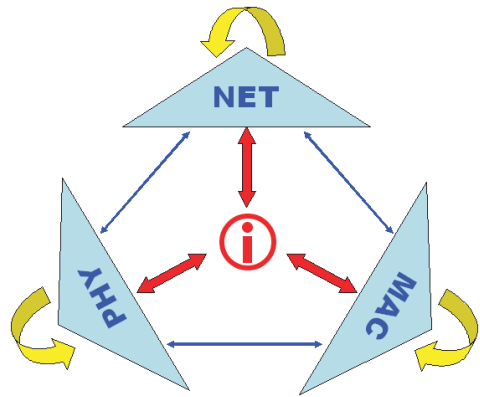}}
\caption{Hierarchical cross-layer design model: interactions amoung Physical, MAC and Network layer.}
\label{fig:cross1}
\end{figure}

Our proposed hierarchical cross-layer design framework seeks to maintain the
advantages of the OSI model, by independently optimizing the interference 
management algorithms based on information sharing among layers. Figure \ref{fig:cross1} illustrates this hierarchical model for the first three layers of the protocol stack: physical layer, MAC (Data Link) Layer and Network Layer. As protocols at different layers act independently to increase 
the energy efficiency in the network, the information exchange between layers 
leads to an iterative adaptation procedure, in which layers take turns to 
adjust and minimize the energy consumption in the network  based on the new interference level and distribution. We note that this hierarchical structure 
raises convergence issues on a vertical plane, and a key issue that should be
addressed is how to appropriately define the information shared between layers,
as well as how to incorporate this information such that the iterative cross-layer adaptation converges, and does not lead to oscillatory behavior. 

In what follows, we propose an energy aware hierarchical joint power 
control and routing design, which we show is guaranteed to converge across 
layers. We then study how further enhancements at the physical layer (i.e., 
multiuser detection receivers in CDMA networks) improve the overall network 
performance. 

\section{Joint Power Control and Routing}

\subsection{Network Model}

We consider an ad hoc network consisting of $N$ mobile nodes. For 
simulation purposes, the nodes are assumed to have a 
uniform stationary distribution over a square area of dimension 
$D^* \times D^*$, but this is not a necessary assumption for the analysis.  
The multiaccess scheme is synchronous direct-sequence CDMA (DS-CDMA) and all
nodes use independent, randomly generated and normalized
spreading sequences of length $L$. The transmitted symbols (assumed to be binary for the purpose of exposition) are detected using
either a matched filter receiver or a linear minimum square error receiver (LMMSE). 
Each terminal $j$ has a transmission power $P_j$ which will be iteratively and 
distributively adapted according to the current network configuration. 
The traffic can be transmitted directly between any two nodes, or it can
be relayed through intermediate nodes. It is assumed that each node generates 
traffic to be transmitted toward a randomly chosen destination node. 
If traffic is relayed by a particular node, the transmissions for different 
sessions at that node are time multiplexed. Also, it is assumed that a 
scheduling scheme is available at the MAC layer to schedule transmission and 
reception minislots for each node. This has the role of avoiding exccesive 
interference between the received and transmitted signals at any particular 
node. The details of the scheduling allocation are beyond the scope of this 
paper. For our design we will use a simplifying worst case assumption that 
will consider that each node creates interference at all times, while in 
reality, some of the time is dedicated only to receiving.
This simplifying assumption supports our hierarchical structure, by avoiding 
interference tracking (routes modification) at the MAC layer time scale.    

We address the problem of meeting Quality of Service (QoS) requirements for
data, i.e., BER (bit error rate) and minimum energy expenditure for the 
information bits transmitted, to conserve battery power. We note that for 
data services, delay is not of primary concern. The target BER requirement 
can be mapped into a target SIR requirement. We note that an optimal target 
SIR can be determined (as in \cite{Goodman:pc}) to minimize the energy per 
bit requirement, under the assumption that data is retransmitted until 
correctly received. 

At a link level, for a given target SIR requirement, the number of 
retransmissions necessary for correct packet reception is characterized by a 
geometric distribution, which depends on the corresponding BER-SIR mapping.
If the transmission rate is fixed for all links, then the energy can be 
minimized by minimizing the transmitted powers on each link. At the physical layer level, this is achieved by power control. However, the achievable minimum powers will depend on the distribution of the interference in the network, and 
thus are influenced by routing. In turn, routing may use power aware  metrics
to minimize the energy consumption. 
The overall cross-layer optimization problem can be formulated as follows.

\begin{equation}
\begin{array}{l} \mbox{minimize} \sum_{i=1}^N P_i \\ \mbox{subject to} \\ \mbox{\hspace{.3cm}} SIR_{(i,j)}({\bf p}) \geq \gamma^*, \ \forall (i,j) \in S^r_a \\
\mbox{\hspace{.3cm}} P_i \geq 0 \\ \mbox{\hspace{-.4 cm} and } r \in {\Large \it \Upsilon}, \end{array}
\label{eq:minp}
\end{equation} 

\noindent where $({\bf p})$ is the vector of all nodes' powers, $S^r_a$ is the set of active links for the current routing 
configuration $r$, obtained using the routing protocol, and $\Upsilon$ is the set of all possible routes.
  
From (\ref{eq:minp}) we can see that optimal power allocation depends on the
current route selection. On the other hand, for a given power allocation,
efficient routing may reduce the interference, thus further decreasing the
required energy-per-bit.
We begin our discussion of the joint optimization of these two effects by 
first considering distributed power control design for a given route 
assignment, which is a classic distributed power control problem for ad hoc 
networks.

\subsection{Distributed Power Control}

\begin{figure}[ht]
\centerline{
\epsfxsize=1.5 in\epsffile{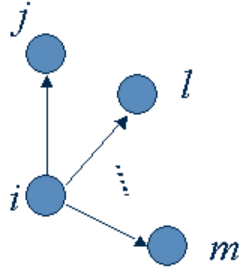}}
\caption{Multiple transmissions from node $i$.}
\label{fig:node}
\end{figure}

In the cellular setting, a minimal power transmission solution is achieved
when all links achieve their target SIRs with equality.  For an ad hoc network,
 implementation complexity constraints may restrict the power control
to adapt power levels for each node, as opposed to optimizing it for each 
active outgoing link for the node.
If multiple active transmission links start at node $i$ 
(Figure \ref{fig:node}), then the worst link must meet the target SIR with
equality. In our model, these outgoing links correspond to destinations for 
various flows relayed by the node, and are used in a time multiplexed fashion.

If we denote the set of all outgoing links from node $i$ as $S^*_i$,
then the minimal power transmission conditions become
\begin{equation}
\min_{k \in S^*_i}SIR_{k} = \gamma^*, \ \forall i =1, 2, ..., N.
\label{eq:qos2}
\end{equation}  

We now express the achievable SIR for an arbitrary active link $(i,j) \in S^r_a$:
\begin{equation}
SIR_{(i,j)}=\frac{h_{(i,j)}P_i}{\displaystyle{\frac{1}{L}\sum_{k=1, k\neq i, k\neq j}^N h_{(k,j)}P_k+ \sigma^2}},
\label{eq:SIR}
\end{equation}  

\noindent where $h_{(i,j)}$ is the link gain for link $(i,j)$, and $\sigma^2$ 
is the background noise power.

Condition (\ref{eq:qos2}) can then be expressed as 
\begin{equation}
\min_{(i,j) \in S^*_i} \frac{h_{(i,j)}P_i}{\displaystyle{\frac{1}{L}\sum_{k=1, k\neq i, k\neq j}^N h_{(k,j)}P_k+ \sigma^2}} = \gamma^*.
\label{eq:condSIR}
\end{equation}  

From (\ref{eq:condSIR}), the powers can be selected as\\

\hspace{3cm} $P_i  = \max_{(i,j)\in S^*_i} \frac{\gamma^*}{h_{(i,j)}}\left[\displaystyle{\frac{1}{L}\sum_{k=1, k\neq i, k\neq j}^N h_{(k,j)}P_k+ \sigma^2}  \right] $
\begin{equation}
\mbox{\hspace{-5.5cm}}= \max_{(i,j)}I_{(i,j)}(\p),
\label{eq:powcond}
\end{equation}  

\noindent where $\p^T = [P_1,\  P_2,\ ..., P_N]$. 

It can easily  be shown that 
$I_{(i,j)}(\p)$ is a standard interference function, i.e., it satisfies the 
three properties of a standard interference function: positivity, monotonicity,
and scalability \cite{yates_fwk}. It was also proved in \cite{yates_fwk} that
$T_i(\p)= \max_{(i,j)}I_{(i,j)}(\p)$ is also a standard interference function.
Since $T_i(\p)$ is a standard interference function, for a feasible system,
an iterative power control algorithm based on 
\begin{equation}
P_i(n+1)=T_i(\p(n)), \ \forall i=1,\ 2, \ ...,\ N,
\label{eq:pc}
\end{equation}

\noindent is convergent to a minimal power solution \cite{yates_fwk}, for 
both synchronous and asynchronous power updates.

Since all the information required for the power updates can be estimated 
locally, the power control algorithm can be implemented distributively.
In particular, a sample average of the square root outputs of the matched 
filter receiver for link $(i,j)$ will determine the quantity 
$E\{y_{(i,j)}^2\}= \frac{1}{L}\sum_{k=1, k\neq i, k\neq j}^N h_{(k,j)}P_k+ h_{(i,j)}P_i+ \sigma^2$.
Further, if the link gain $h_{(i,j)}$ is also estimated, all information 
required  for power updates at node $i$ is available locally.

\subsection{Joint Power Control and Routing}

The previous subsection has proposed an optimal power control algorithm,
which minimizes the total transmitted power given SIR constraints for all
active links, for a given network configuration.
However, the performance can be further improved by optimally choosing the
routes as well. Finding the optimal routes to minimize
the total transmission power over all possible configurations is an NP-hard
problem. 

\begin{figure}[ht]
\centerline{
\epsfxsize=2.7 in\epsffile{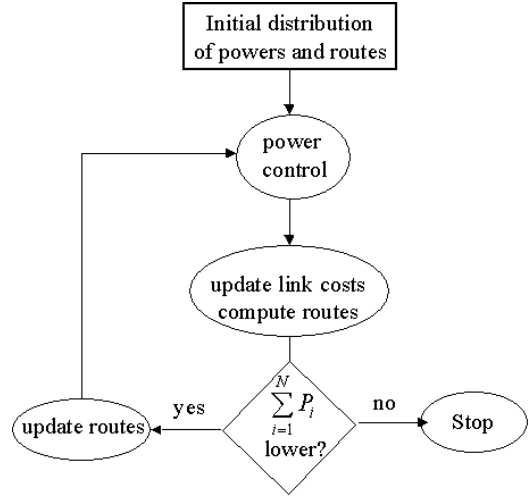}}
\caption{Joint power control and routing}
\label{fig:joint}
\end{figure}

We propose a suboptimal solution, based on iterative power control and routing,
which is shown to converge rapidly  to a local minimum energy solution.
This solution is compatible with our proposed hierarchical cross-layer framework, by promoting independent protocol updates with information sharing accross 
layers. More specifically, we propose a joint algorithm, that alternates 
between power control (at the physical layer) and route assignments (at the network layer), until further improvements in the 
energy consumption cannot be achieved. At each step of the algorithm, 
the power control optimizes powers based on the current route assignment, while
after power assignment, new minimum energy routes are determined based on
the current power distribution of the nodes (see Figure \ref{fig:joint}). 

As we have mentioned in Section IIIA, the optimization problem that we are 
solving can be expressed as in (\ref{eq:minp}), i.e., we try to minimize the 
sum of transmission powers, subject to SIR constraints, by both power control 
and route assignments. We note that the target SIR requirement is selected 
such that a BER requirement is met for a fixed prescribed rate allocation, 
determined by a prescribed spreading gain. Thus, in our system model the 
transmission rate is fixed.  
 
In the previous section, we have described how the transmission powers are 
chosen for each node given a current route configuration, and we have shown that, for our system model, they are 
unique per node, no matter which flow is currently relayed by the node. 

Thus, the information that the network layer sees is the vector of powers for 
all the nodes, $\p^T=[P_1, \ P_2, \ ..., P_N]$, which completely characterizes 
the interference distribution in the system, given a certain location for the 
nodes.  

For routing, we use Dijkstra's algorithm  \cite{Berts_dn}, \cite{BNT} with
associated costs for the links. In order to try to minimize further the total transmitted power in the network, a natural choice of costs for the routing, would be based on the transmission power spent by a node sending on a given link. However, for convergence reasons for the cross-layer algorithm (which will be explained shortly), the cost for an arbitrary link $(i,j)$ is 
determined as

\begin{equation}
c(i,j)=\left\{\begin{array}{l} P_i  \mbox{\hspace{2cm} if } SIR_{(i,j)}\geq\gamma^* \\
\infty  \mbox{\hspace{2cm} if } SIR_{(i,j)}< \gamma^* \end{array} \right..
\label{eq:lc}
\end{equation}

The reason for choosing the link costs as in (\ref{eq:lc}) is that we would 
like to restrict the pool of links available for routing to include only links
that already meet the target SIR. As we will see shortly, this condition will
ensure the convergence of the algorithm towards a minimum energy solution.

To determine a better possible routing option, we need to evaluate the new costs for all links, given the current distribution of powers resulted from the
previous power control step. In order to determine the routing costs for the links that are not currently active, the 
achievable SIR for these links must be estimated. This requires that each node
$i$ update a routing table which should contain the estimated link gains 
toward all the other nodes, $h_{(i,j)}$, $j=1,2, ..., N$, $j \neq i$, the
transmitted powers of all nodes, $P_j$, $j=1,2, ..., N$, and the extended estimated interference at 
all the other nodes, defined as $\tilde{I}(i,j) = \sum_{k=1, k\neq i, k\neq j}^N h_{(k,j)}P_k + h_{(i,j)}P_i$,  $j=1,2, ..., N$, $j \neq i$. Hence, the estimated SIR for link
$(i,j)$ can be expressed as
\begin{equation}
\widetilde{SIR}_{(i,j)}=\frac{h_{(i,j)}P_i}{\displaystyle{\frac{1}{L}\left(\tilde{I}(i,j)-h_{(i,j)}P_i \right) + \sigma^2}}.
\label{eq:SIRest}
\end{equation}  

We note that the achievable SIR on any potential link (currently active or not)
depends only on the current distribution of nodes, and on the current power
assignment, and does not depend on the current assigned routes, and 
consequently does not change for new route assignments. This property is a 
result of the fact that multiple sessions are time multiplexed at a node, and
are all transmitted with the same power, such that the transmitted power for a node $i$ is fixed and equal to $P_i$. This result can be summarized in the
following proposition.

{\prop For a given distribution of nodes in the network, after the convergence of the power control algorithm, the achievable SIR on any arbitray link,
depends only on the nodes' transmitted powers and is independent of the
current route assignment.
\label{prop:SIR}
}
 
We note that if sessions are not time multiplexed at a relaying node, the 
above proposition does not hold any more (e.g. the total power transmitted by a node is additive over the number of relayed flows for multi-code transmission, and thus depends on the routing configuration), and the convergence of the proposed 
joint-power control algorithm is not guaranteed.
However, as a disadvantage for the time multiplexed scheme, the throughput per
session is limited by the number of sessions relayed by a node. In an extension
of this work \cite{ciss_nie04}, we also have proposed a cost modification for the 
routing to account for this effect, which yielded a more uniform 
distribution of relayed flows per node over the entire network. Also, in 
\cite{vtc_hasan05}, we have compared the performance of a time multiplexed scheme, with 
the case in which multi-code CDMA is used for simultaneous transmission of all relayed flows (which increases the interference in the system).

Starting from an initial distribution of powers and routes, and assuming that
the system is feasible for the initial configuration, the joint power control 
and routing algorithm is summarized in Figure \ref{fig:joint}. 

{\theorem For a feasible initial network configuration, the joint power control 
and routing algorithm converges to a locally minimal transmitted power solution.
\label{prop:jointpc}}

{\it {Proof:}} As we previously showed, for a feasible initial network 
configuration, the power control minimizes the total transmitted power, 
while ensuring that all active links meet their SIR requirements: 
$SIR_{(i,j)} \geq \gamma^*$, $\forall (i,j) \in S_a^r$. After the convergence 
of the power control algorithm, the link costs are estimated and updated 
according to (\ref{eq:lc}) and (\ref{eq:SIRest}), and a minimal cost route, 
equivalent to a minimal transmitted power route, is selected for each session. 
As a consequence, the new routes are selected such that the sum of all 
transmitted powers for all active links is minimized, while the SIR 
constraints are met for all links (from Proposition 
\ref{prop:SIR} and (\ref{eq:lc})). If no power improvements can be achieved,
the algorithm stops. Otherwise, the sum of transmission powers decreases after
the route selection. Since all the new active links satisfy $SIR_{(i,j)} 
\geq \gamma^*$, $\forall (i,j) \in S_a^r$, the system is feasible, and 
therefore, the power control algorithm produces a decreasing sequence of 
power vectors converging to a minimal power solution \cite{yates_fwk}.

Hence, each step of the iteration (power control or routing) produces an
improvement in the total transmitted power, while meeting SIR requirements
for all active links. The algorithm stops at a locally minimal transmitted 
power solution, where no further decrease in transmission power can be achieved
by the routing protocol.$\hfill$ $\Box$ 

We note that the locally minimal transmitted power solution achieved by the 
proposed algorithm depends on the initial network configuration chosen.
For initialization, we propose an algorithm similar to that which was proposed 
in \cite{admud}. We first select an initial
distribution of powers (equal powers or random distribution) and then determine
routes by assigning link costs equal to the energy per bit consumption defined in (\ref{eq:eb_def}). This approach also permits us to quantify the energy 
requirement improvements of the joint optimization relative to the initial starting point.

We note that the total energy requirement depends on the
current initialization for the powers. To improve the expanded energy with
minimal complexity increase, the algorithm can be run several times with 
different random power initializations, and the best energy solution over all 
runs can be determined.
   
\subsection{Simulations}

In this section, we present some numerical examples for ad hoc networks 
with 55 and 40 nodes, respectively, uniformly distributed over a square area 
of $200 \times 200$ meters. The target SIR is selected to be 
$\gamma^* = 12.5$ (which was shown to be an optimal value that minimizes energy per bit consumption for an FSK scheme \cite{Goodman:pc}), and
the noise power is $\sigma^2 = 10^{-13}$, which approximately 
corresponds to the thermal noise power for a bandwidth of 1 MHz. We consider 
low rate data users, using a spreading gain of $L=128$.
For this particular example, we choose equal initial transmit powers, 70 dB 
above the noise floor ($P_t = 10^{-6}$W), and a path loss model with path 
loss coefficient $n=2$.

\begin{figure}[ht]
\centerline{
\epsfxsize=2.7 in\epsffile{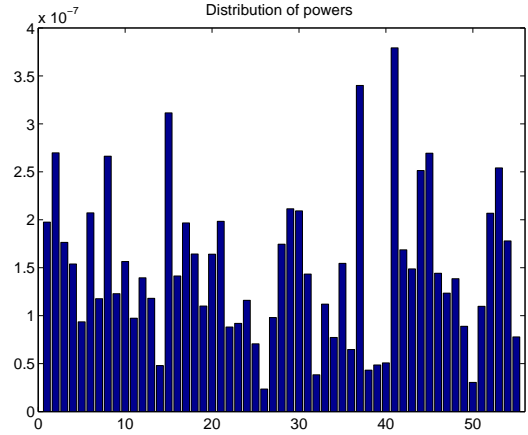}}
\caption{Distribution of powers after convergence}
\label{fig:pows2}
\end{figure}

\begin{figure}[ht]
\centerline{
\epsfxsize=2.7 in\epsffile{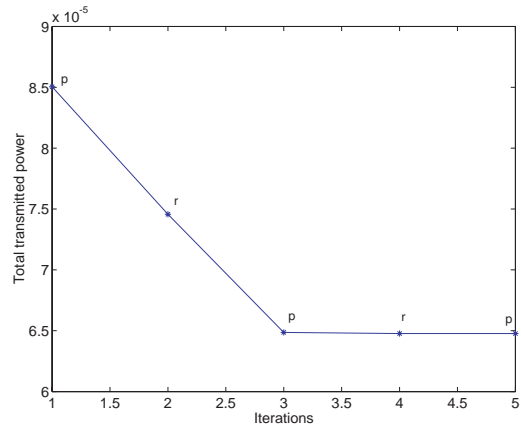}}
\caption{Total transmission power}
\label{fig:p_net1}
\end{figure}

\begin{figure}[ht]
\centerline{
\epsfxsize=2.7 in\epsffile{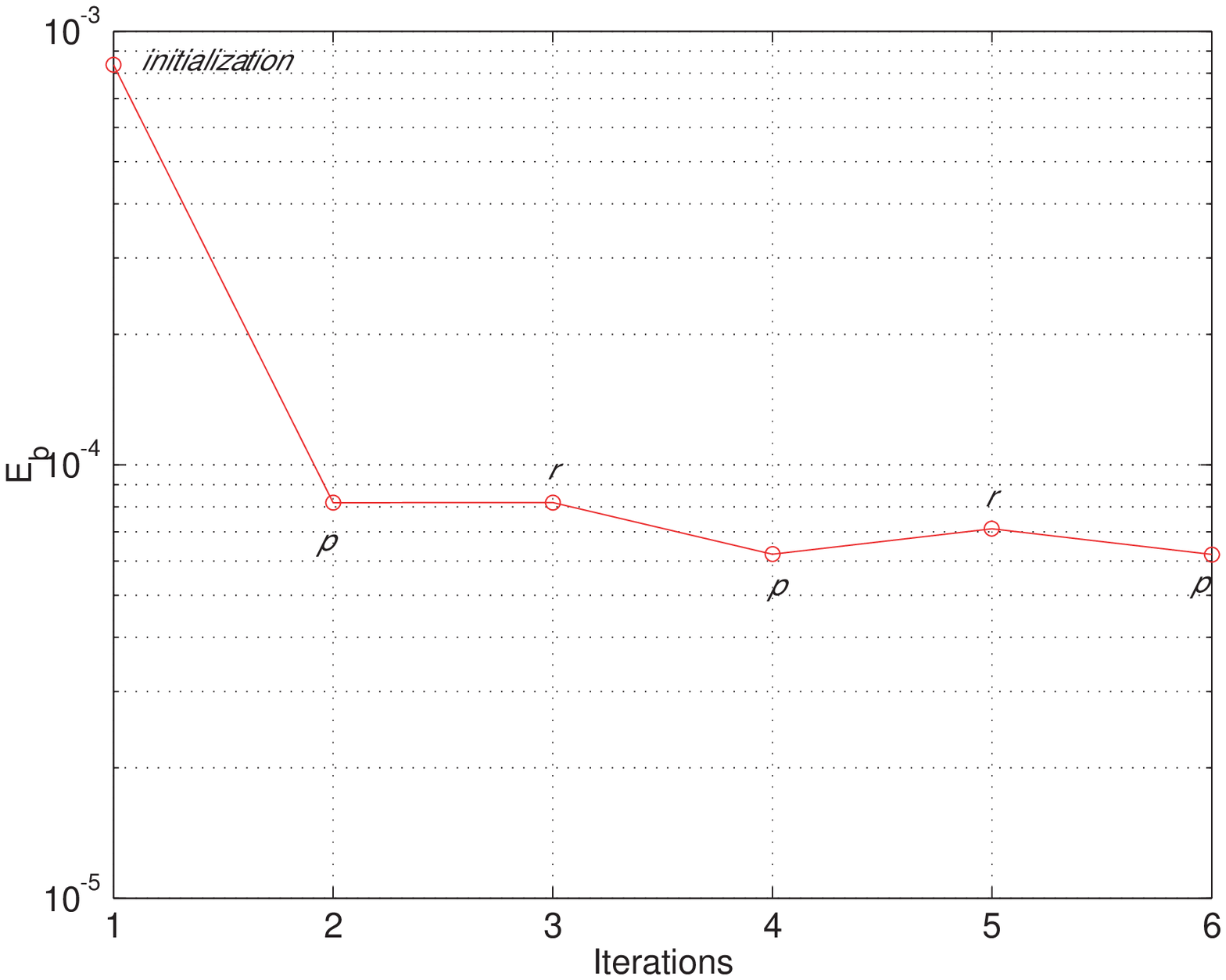}}
\caption{Energy per bit}
\label{fig:e_net1}
\end{figure}

In Figure \ref{fig:pows2} we show the final  
distribution of powers after the convergence of the joint power control and
routing algorithm.   
Figures \ref{fig:p_net1} and \ref{fig:e_net1} illustrate the performance 
of the proposed joint optimization algorithm. In Figure \ref{fig:p_net1}, 
it can be seen that the total transmitted power in the network progressively 
decreases as the proposed algorithm iteratively optimizes power and routes.
The values in Figure \ref{fig:p_net1} represent the total transmitted power 
obtained over a sequence of iterations:
[power control, routing, power control, routing, power control].
In Figure \ref{fig:e_net1}, the achieved energy-per-bit is compared for the
same experiment with the first energy value, which represents the 
energy-per-bit obtained in the initial state. It can be seen that substantial
improvements are achieved by the proposed joint optimization algorithm.

Note that, at the end of each iteration pair
[routing, power control], the energy is further minimized. However, after
new routes are selected, the powers are not yet optimized, so it is possible
that previous routes might have better energy-per-bit performance (for the
same power allocation, higher SIRs may improve the energy consumption).
 
\begin{figure}[ht]
\centerline{
\epsfxsize=2.7 in\epsffile{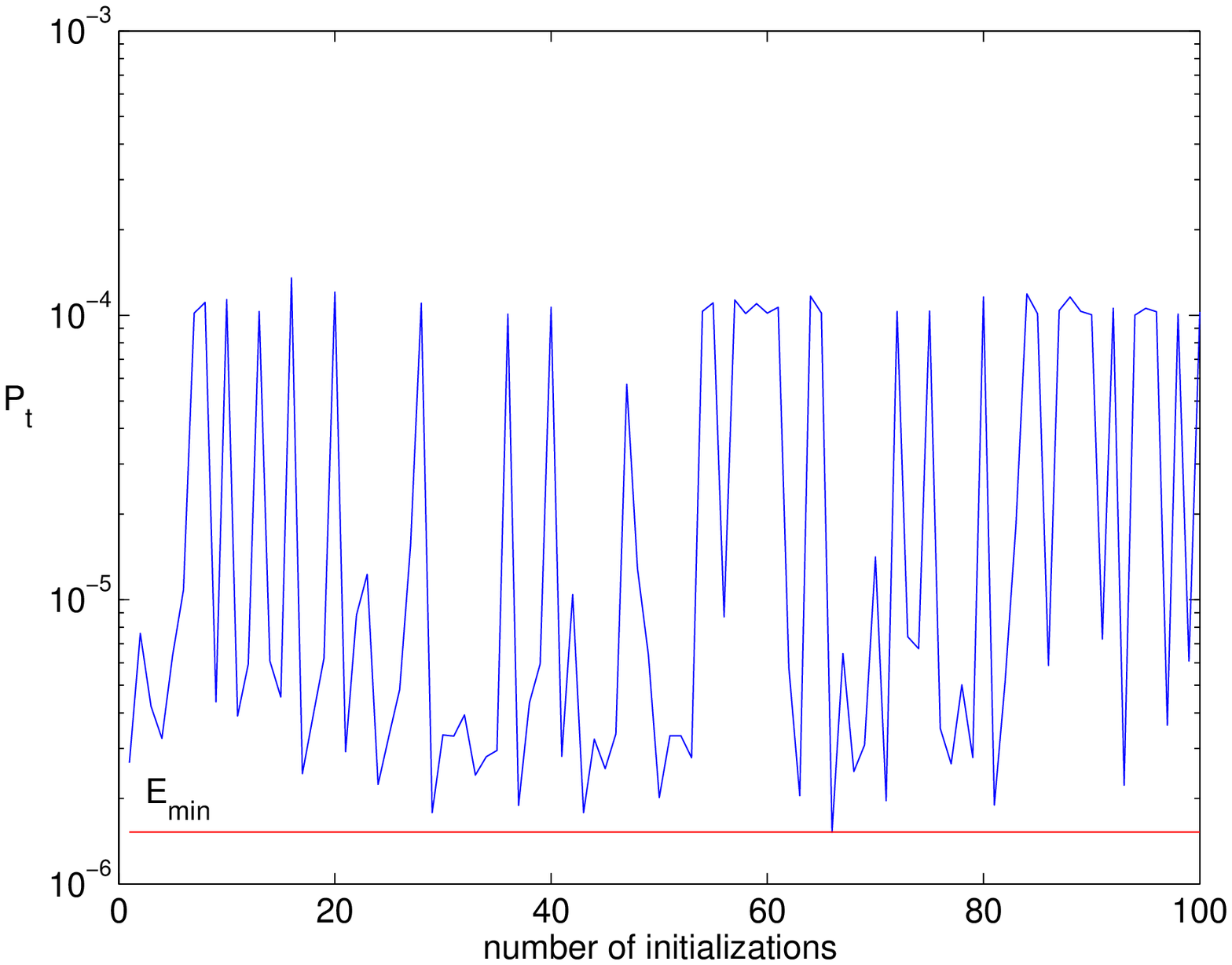}}
\caption{Energy function for different initializations}
\label{fig:energ}
\end{figure}

\begin{figure}[ht]
\centerline{
\epsfxsize=2.7 in\epsffile{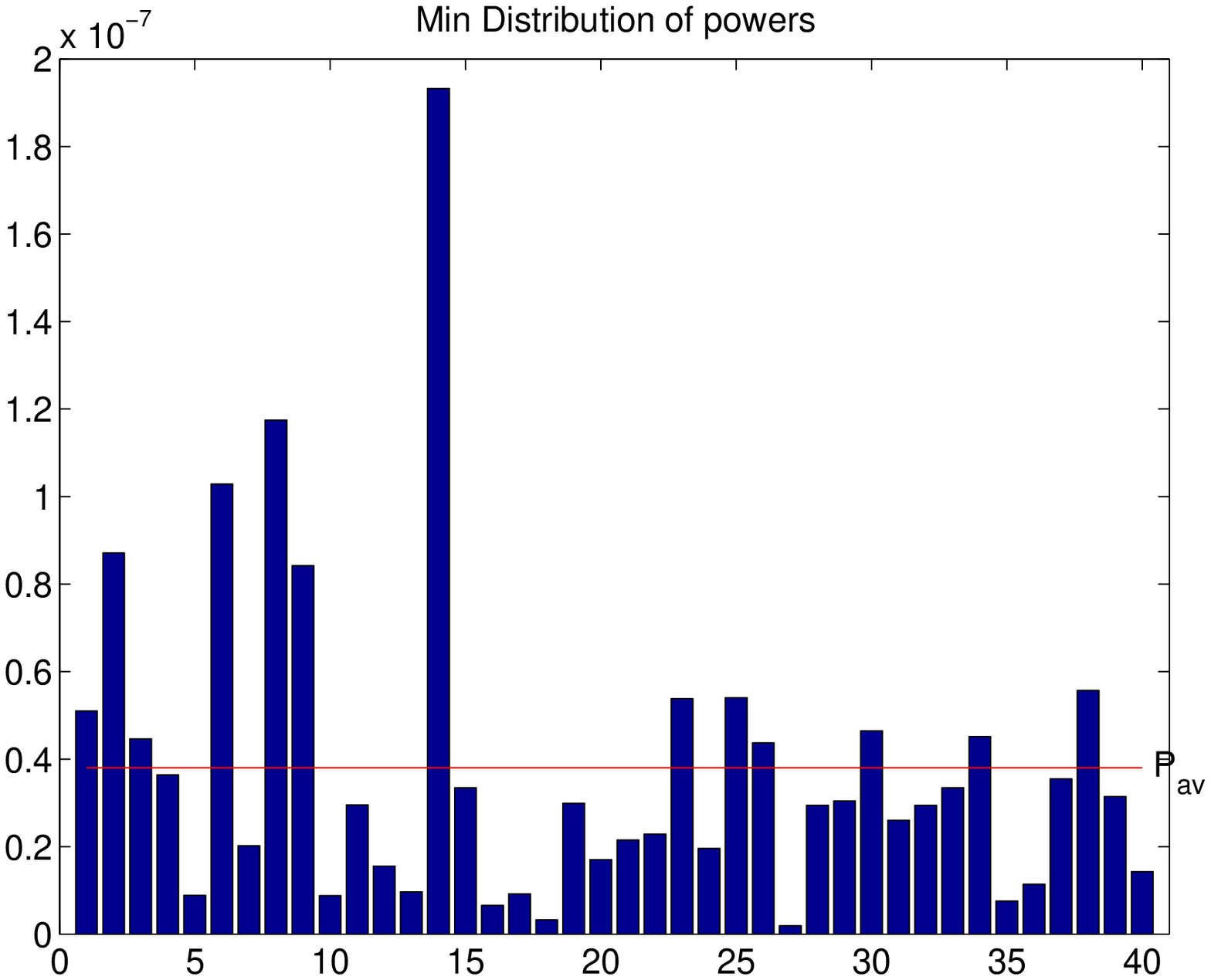}}
\caption{Distribution of powers for the minimal energy solution}
\label{fig:minrun}
\end{figure}

As we have previously mentioned, the actual energy results after convergence
depend on the initial starting point for the algorithm. In Figure 
\ref{fig:energ}, we illustrate the variation in the total transmission power
obtained with various initializations (100 trials are considered) for an 
ad hoc network with 40 nodes. We can see that significant energy improvements 
can be achieved if the algorithm is run repeatedly with different 
initializations and the best configuration is selected. In Figure 
\ref{fig:minrun} we show the final distribution of powers for this minimal
energy solution.

\subsection{Uniform energy consumption}

While we saw that the power distribution in Figure \ref{fig:minrun} gives a very 
low total energy consumption, this solution leads to unequal power consumption
among nodes, which ultimately results in shorter life span for certain 
nodes (e.g. node 14 in Figure \ref{fig:minrun}). Note that in mobile
nodes, this problem is overcome by the fact that node locations change with
time, so in the long run, the power consumption tends to be more uniform. 

For fixed nodes, or slow moving ones, we overcome this problem by selecting
 a set of alternate ``good routes'' ($N_s$ routes) 
and their corresponding power distributions. The routes (and power vectors)
are then randomly assigned, such that the power consumption variance among 
nodes is minimized. 
A route $i$ and its corresponding power vector 
$\p_i$ are selected from the initial set of ``good routes'', with probability
$w_i$. The probabilities $w_i$, $i=1,\ldots, N_s$ are assigned to routes
such that the following conditions hold
\[\min_{\w}\parallel \Pp-P_{av} \parallel^2_2;\]
\begin{equation}
\left\{ \begin{array}{l} 0 \leq w_i \leq 1, \ i=1 \ldots, N_s; \\ \sum_{j=1}^{N_s}w_j =1, \end{array} \right.
\end{equation} 

\noindent where $\w = [w_1, \ w_2, \ldots, w_{N_s}]$, $\Pp = [\p_1, \ p_2, \ldots \p_{N_s}]$, and $P_{av}$ is the average power consumption across nodes obtained
for the minimal energy solution.

\begin{figure}[ht]
\centerline{
\epsfxsize=2.7 in\epsffile{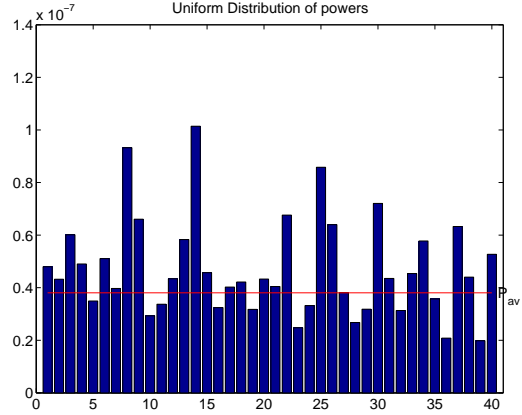}}
\caption{Energy per bit}
\label{fig:unifrun}
\end{figure}

Alternatively, routes can be assigned deterministically, such that $w_i$ 
represents the fraction of time route $i$ and its corresponding power vector
are selected for transmission. In Figure \ref{fig:unifrun} we illustrate how
the power distribution changes in the ad hoc network when $N_s=9$ ``good routes''  are selected. 
These routes (and their corresponding power distribution) 
are selected to be within $10\%$ of the minimal energy solution obtained with 
100 different random initializations. Comparing the results from Figure 
\ref{fig:unifrun} with the ones in Figure \ref{fig:minrun}, we can see a more
uniform consumption across all nodes in the ad hoc network.

\section{Joint Power Control, Routing and Multiuser Detection}

To extend the above described joint power control and routing algorithm to include
receiver optimization we build on results on iterative, distributed, joint power control
and minimum mean square error multiuser detection presented in \cite{pc-mmse}.
In \cite{pc-mmse} an iterative two-step integrated power control and multiuser detection
algorithm was proposed, for which, in the first step, the LMMSE filter coefficients are adjusted
according to the current vector of powers $\p$ (equation \ref{eq:LMMSE_filt}), then in  
the second step, a new power vector is selected for the given filter coefficients:

\begin{itemize}
\item Step 1: Optimize filter coefficients given the power vector $\p^T = [P_1, P_2, \cdot, P_n]$:
\begin{equation}
\hat{\ci} = \frac{\sqrt{P_i(n)}}{1+P_i(n)\si^T \A_i^{-1}(\p(n))\si}A_i^{-1}(\p(n))\si,
\label{eq:LMMSE_filt}
\end{equation}
\noindent where $\ci$ and $\si$ are the filter coefficients vector, and the signature sequence vector for user $i$, 
respectively, $n$ is the iteration number, and $\A_i$ is defined as $\A_i = \sum_{j\neq i}P_j h_{ij}\sj\sj^T$. \\

\item Step 2: Optimize powers based on currently selected filter coefficients:
\begin{equation}
P_i(n+1)=\frac{\gamma_i^*}{h_{ii}}\frac{\sum_{j \neq i} P_j(n)h_{i,j}(\hat{\ci}^T\sj)^2+\sigma^2\hat{\ci}^T\hat{\ci}}{(\hat{\ci}^T\si)^2}.
\label{eq:power_sel}
\end{equation}

\end{itemize}

Given the above algorithm, to extend our joint power control and routing scheme to include
receiver optimization, we simply replace the simple power control adaptation at the 
physical layer, by the above joint power control and multiuser detection algorithm.

\begin{figure}[ht]  
\centerline{
\epsfxsize=2.8in\epsffile{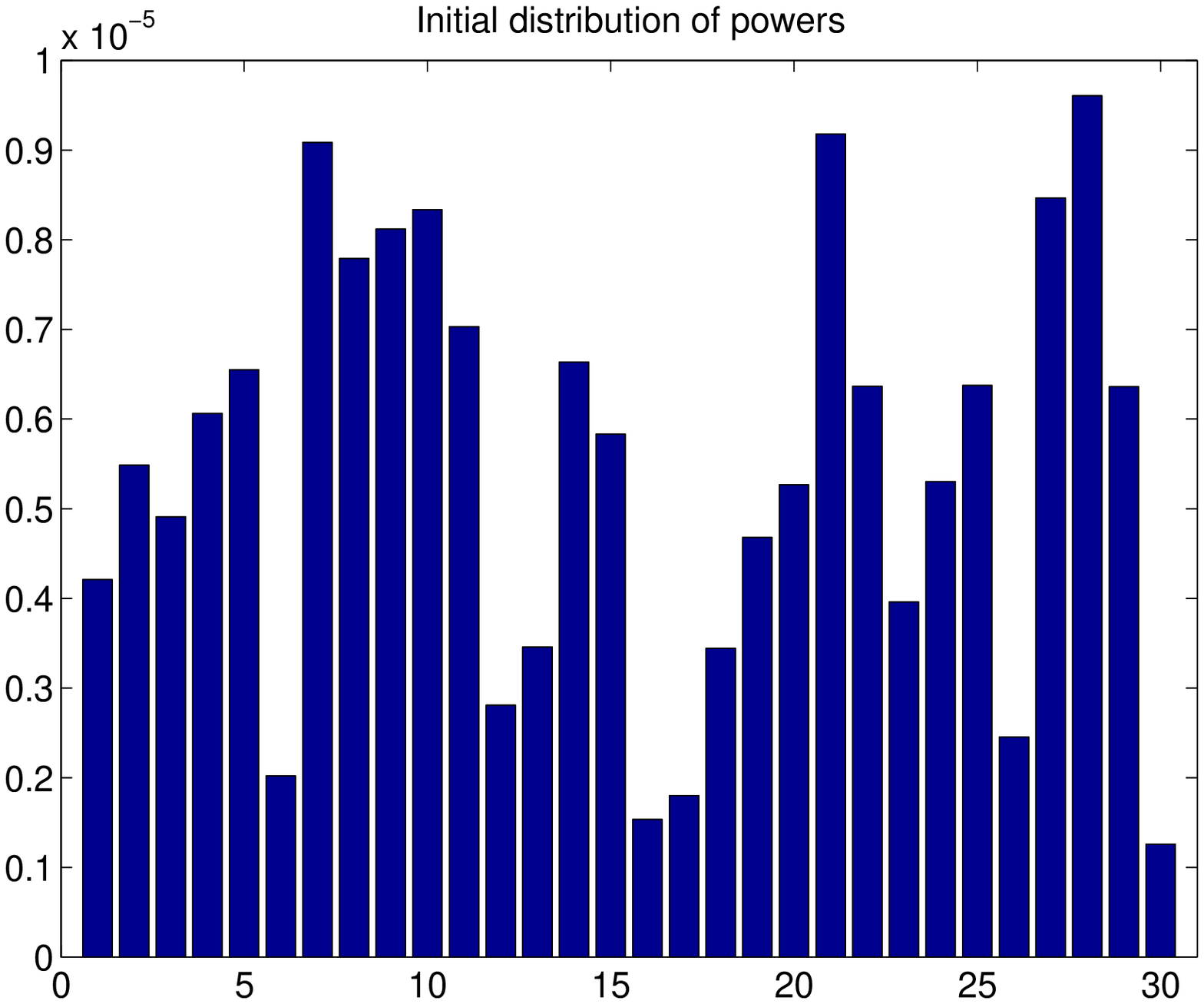}\hspace{.8pc}}

\hspace*{2.3in}(a)\\
\centerline{
\epsfxsize=2.8in\epsffile{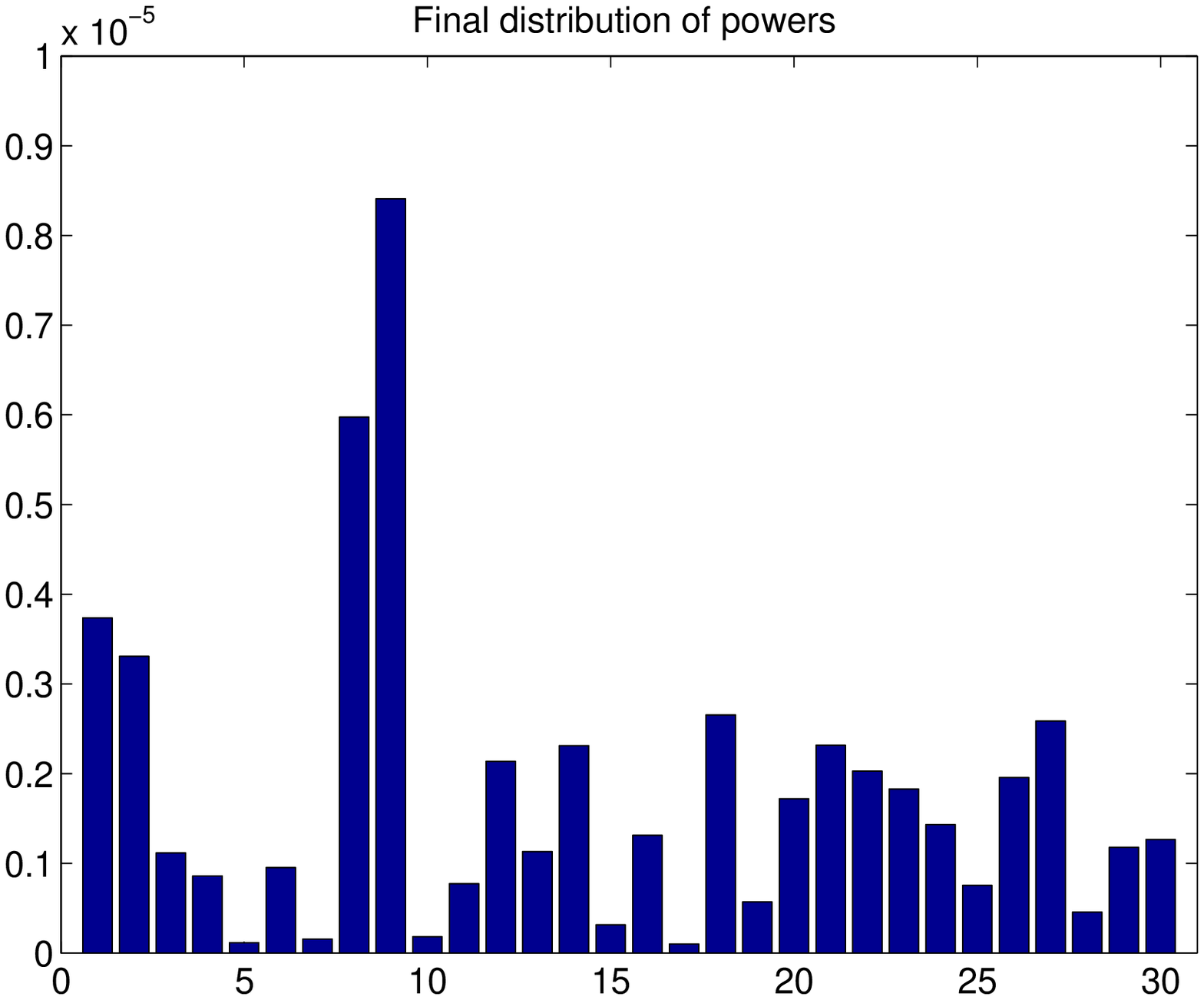}\hspace{.8pc}} 

\hspace*{2.3in}(b)
\caption{Joint power control, multiuser detection and routing: Distribution of powers versus node number, (a) initially, (b) after convergence}
\label{fig:pows}
\end{figure}

\begin{figure}[ht]
\centerline{
\epsfxsize=2.7 in\epsffile{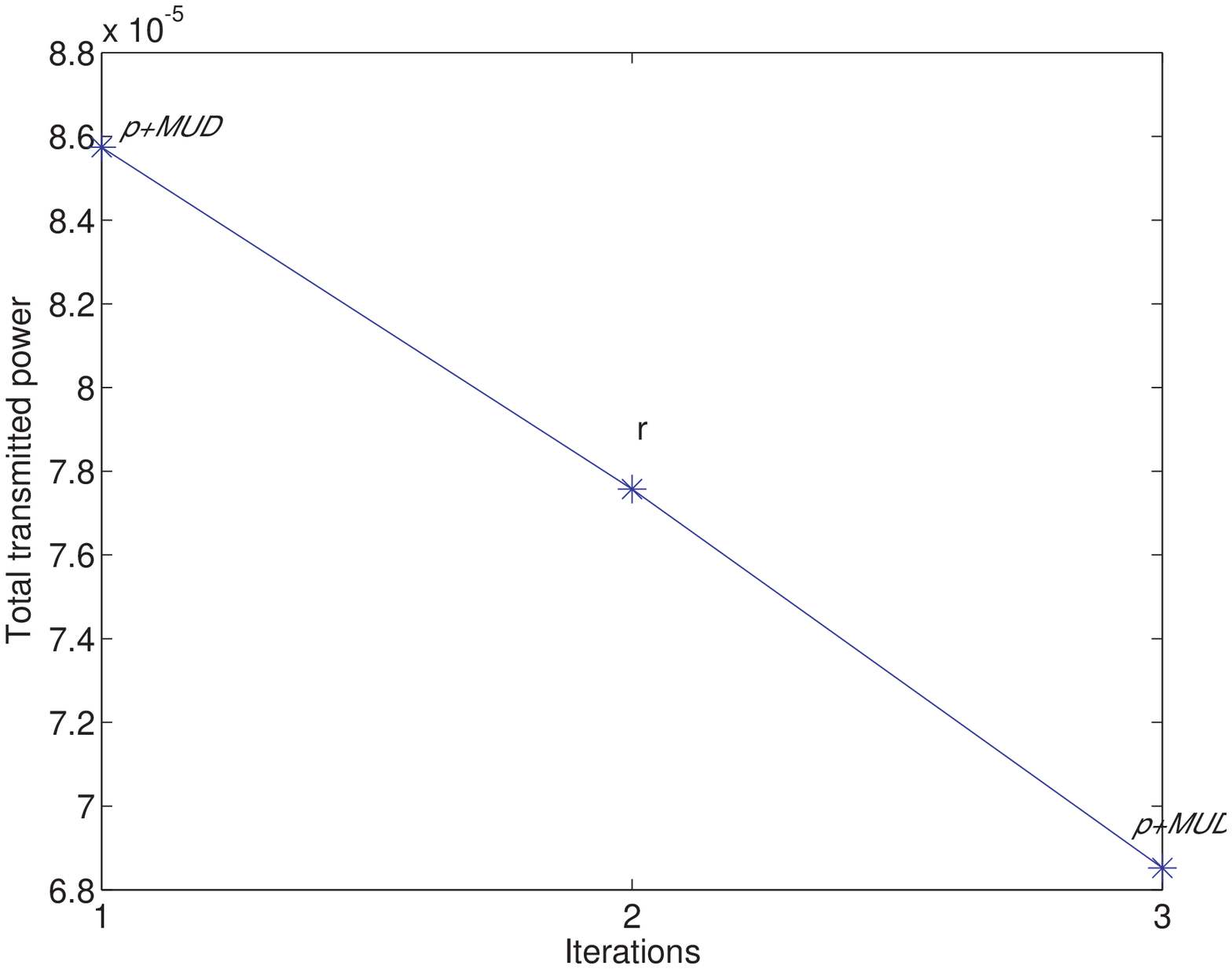}}
\caption{Total transmission power: joint power control, multiuser detection and routing}
\label{fig:p_net}
\end{figure}

\begin{figure}[ht]
\centerline{
\epsfxsize=2.8 in\epsffile{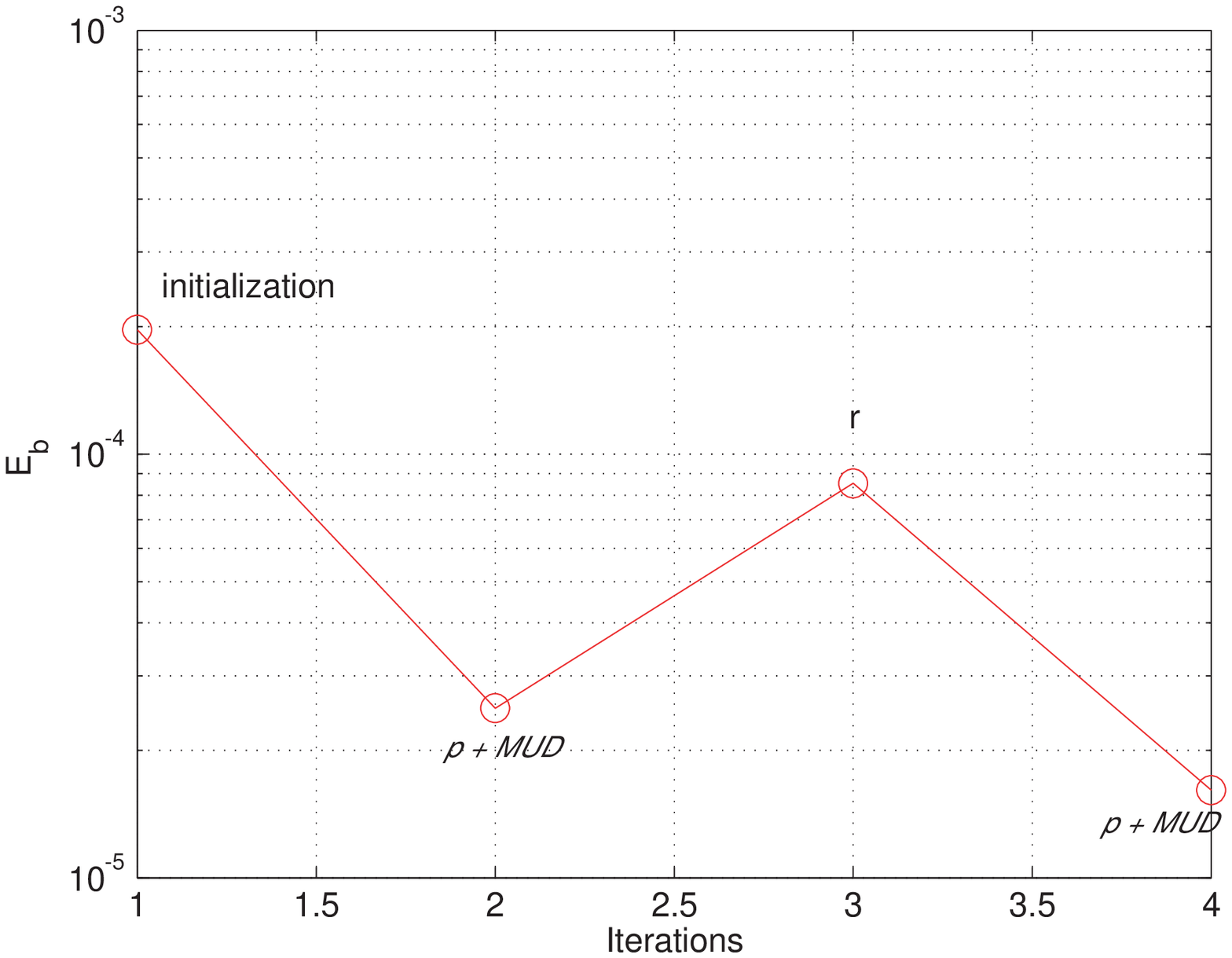}}
\caption{Total energy consumption: joint power control, multiuser detection and routing}
\label{fig:e_net}
\end{figure}

Simulation results show a very similar convergence behavior and energy savings for the
joint power control, multiuser detection and routing algorithm, compared to the solution with
matched filters (see Figures \ref{fig:pows}, \ref{fig:p_net} and \ref{fig:e_net}). 
We also note a significant capacity increase when multiuser detection is 
employed. We use as a capacity measure the total throughput that can be supported by the network
such that the power control is feasible for a target SIR of $\gamma^* = 12.5$. 
We note that the power control feasibility depends on the actual network topology. To determine
the maximum load for the network, we randomly generated 100 different topologies (for the same
number of users) and we selected the maximum number of users (for a given spreading gain) 
that yielded  feasible topologies $95$\% of the time, for a given initial power
distribution for the nodes.

For the matched filter case, we selected $L = 128$ and the maximum number of users that met the 
feasibility condition was determined to be $N=55$. For the LMMSE case, since the capacity increases
significantly, to reduce the complexity of the simulation (the number of nodes), we have selected
$L=32$, with a  resulting capacity of $N=30$. This yielded a total normalized throughput gain 
for the LMMSE case of

\begin{equation}
T_{g(LMMSE)}=\frac{N_{LMMSE} \times L_{MF}}{L_{LMMSE}\times N_{MF}} = 2.18.
\end{equation}    

To illustrate the performance of the joint power control, multiuser detection and routing protocol, 
we have considered similar network parameters as before, with the sole difference of selecting
$N=30$ and $L=32$. Random initial transmission powers were selected, approximately 70 dB 
above the  {noise floor}.

Figure \ref{fig:pows}
shows the initial distribution of powers, as well as the optimal power control
distribution after convergence.   

Figures \ref{fig:p_net} and \ref{fig:e_net} illustrate the performance 
of the proposed joint optimization algorithm with multiuser detection. In Fig. \ref{fig:p_net}, 
it can be seen that the total transmitted power in the network progressively 
decreases as the proposed algorithm iteratively optimizes power, filter coefficients, and routes.
The values in Fig. \ref{fig:p_net} represent the total transmitted power 
obtained over a sequence of iterations:
[power control + MUD ,  {routing}, power control+ MUD ,  {routing}, power control+MUD].

In Fig. \ref{fig:e_net}, the achieved energy-per-bit is compared for the
same experiment with the initial energy value (with randomly selected powers). 
It can be seen that substantial improvements are achieved by the proposed 
joint optimization algorithm (approximately one order of magnitude).

\section{Conclusions}
In this paper, we have proposed joint power control and routing 
optimization for wireless  ad hoc data networks with energy constraints.
Both energy minimization and network lifetime maximization have been considered as optimization
criteria. We have shown that energy savings of an order of magnitude can be obtained,
compared with a fixed transmission power, energy aware routing scheme.
Our proposed algorithm is based on a hierarchical cross-layer framework which maintains
the advantages of the OSI layered architecture, while allowing for protocol optimization
based on information sharing between layers.
The network capacity has been further enhanced by employing multiuser detection, with 
a similar obtained energy performance.
Our simulation results show that our distributive joint
optimization algorithm converges rapidly towards a local minimum
energy. The rapid convergence of the power-routing 
protocol makes it suitable for implementation in mobile ad hoc networks.

\bibliography{ref_adhoc}
\bibliographystyle{abbrv}

\end{document}